\def\aa{{A\&A}}

\def\annrev{{ARA\&A}}
\def\apj{{ApJ}}

\def\mnras{{MNRAS}}
\def\nat{{Nature}}

\documentclass[cup5b]{caps}

\usepackage{graphicx}
\usepackage{amssymb}
\usepackage{ociwsymp1e}           
\HeadText{T. Alexander}

\begin{document}

\newcommand{\Ua}{\widetilde{a}}
\newcommand{\UP}{\widetilde{P}}
\newcommand{\Um}{\widetilde{m}}
\newcommand{\Ut}{\widetilde{t}}
\newcommand{\Ur}{\widetilde{r}}
\newcommand{\Urp}{\widetilde{r}_{p}}
\newcommand{\Urs}{\widetilde{r}_{s}}
\newcommand{\UdE}{\Delta \widetilde{E}}
\newcommand{\Urt}{\widetilde{r}_{t}}
\newcommand{\Urh}{\widetilde{r}_{h}}
\newcommand{\Us}{\widetilde{\sigma}}
\newcommand{\UE}{\widetilde{E}}
\newcommand{\UR}{\widetilde{R}}
\newcommand{\Mo}{M_{\odot }}
\newcommand{\Ro}{R_{\odot }}
\newcommand{\Lo}{L_{\odot }}
\newcommand{\HLm}{\widehat{L}_{0}}
\newcommand{\Utm}{\widetilde{t}_{0}}
\newcommand{\UT}{\widehat{T}}
\newcommand{\ULt}{\widetilde{L}_{t}}
\newcommand{\HLt}{\widehat{L}_{t}}
\newcommand{\Utd}{\widetilde{\tau }_{d}}
\newcommand{\Ms}{M_{\star }}
\newcommand{\Rs}{R_{\star }}
\newcommand{\Ls}{L_{\star }}
\newcommand{\Ts}{T_{\star }}
\newcommand{\Ns}{N_{\star }}
\newcommand{\UWp}{\widetilde{\Omega }_{p}}
\newcommand{\UdW}{\Delta\widetilde{\Omega }}
\newcommand{\UdJ}{\Delta \widetilde{J}}
\newcommand{\UI}{\widetilde{I}}
\newcommand{\Te}{T_{\mathrm{eff}}}
\newcommand{\en}{\overline{n}}

\pagenumbering{arabic}

\setcounter{chapter}{1}

\author[]{T. ALEXANDER\\The Weizmann Institute of Science, Rehovot, Israel}
\chapter*{Stellar Tidal Processes Near \\ Massive Black Holes\\
\vspace{1 em}
{\rm\normalsize T. ALEXANDER\\[-1em]
 \em The Weizmann Institute of Science, Rehovot, Israel}
}
\thispagestyle{ociwfirstheader}

\begin{abstract}
Close tidal interactions of stars with a central massive black hole (MBH)
or with other stars in the high density cusp around it can affect a
significant fraction of the stellar population within the MBH radius of
influence. We consider three strong processes that have the potential of
modifying stellar structure and evolution there. (1) Tidal spin-up by
hyperbolic star-star encounters. (2) Tidal scattering of stars on the
MBH. (3) Tidal heating of inspiraling stars---``squeezars''---that were
tidally captured by the MBH. We discuss the implications for stellar
populations near MBHs and for the growth of MBHs by tidal disruption of
stars, and the possible observational signature of such processes near the
MBH in the Galactic Center. We compare the event rates of prompt tidal
encounters (tidal disruption and tidal scattering) and slow inspiral events
(squeezars / tidal capture), and find that tidal capture is at least an
order of magnitude less efficient than prompt disruption. This means that
past studies, which assigned similar weights to prompt disruption and tidal
capture, over-estimated the contribution of tidal disruption to the growth
of the MBH by at least a factor of two.
\end{abstract}

\section{Introduction}
\label{s:intro}

Strong tidal interactions involving stars are expected to occur frequently
near a MBH in a galactic center. 

First, the MBH is a mass sink, which drives a flow of stars from the MBH
radius of influence $r_h$ to the center, to replace those it has destroyed. An
inevitable consequence of this flow is that some stars are deflected into
orbits whose periapse $r_p$ lies just outside the critical radius for
destruction. We will focus here on the case where the MBH mass $m$ is small
enough so that the tidal disruption radius $r_t\!\sim\!  \Rs(m/\Ms)^{1/3}$,
where $\Ms$ and $\Rs$ are the stellar mass and radius, lies outside the
event horizon $r_s$, ($m\!\lesssim\!10^8 \Mo$ for a solar type star). Such
stars will suffer an extreme tidal impulse, but will not be destroyed, at
least not on their first peri-passage. There are two possible outcomes:
that the star is ultimately disrupted, or that it avoids subsequent
encounters with the MBH. Both are considered in detail below.

Second, a variety of formation scenarios predict that MBHs should lie in
the center of a high density stellar cusp (e.g. Bahcall \& Wolf 1977; Young
1980). The diverging stellar density implies that there must be some volume
around the MBH where close tidal encounters occur on timescales
significantly shorter than the typical stellar lifetime. Such encounters
will have a very different outcome from those that occur in globular
clusters that do not contain a MBH. In most cases the
encounters will not lead to tidal capture. Instead the two stars will
continue on their separate ways after experiencing a brief strong tidal
impulse.

Extreme tidal interactions, which transfer energy and angular momentum from
the orbit to the star, can affect its structure and subsequent evolution by
heating it, spinning it up, mixing it, or ejecting some of its mass. This
is interesting in view of the observed presence of unusual stellar
populations near MBHs: the blue nuclear cluster in the inner $\sim\!0.02$
pc of the GC (Genzel et al. 1997), and around the MBH in M31 (Lauer et
al. 1998); evidence for anomalously strong rotational dredge-up in an M
supergiant near the MBH in the GC (Carr, Sellgren \& Balachandran 2000),
but not in a high density nuclear cluster without a MBH (Ram\'{\i}rez et
al. 2000); the unusually high concentration of very rare extreme blue He
supergiants around the Galactic MBH (Krabbe et al. 1991; Najarro et
al. 1994).

The observational consequences of extreme tidal interactions cannot be
predicted with certainty at this time, although some reasonable conjectures
can be made (Alexander \& livio 2001; Alexander \& Morris
2003). Irrespective of this uncertainty, it is clear, as shown below, that
the amount of tidal energy deposited in the star can reach a significant
fraction of its binding energy, and that the angular momentum extracted
from the orbit can spin a star up to a significant fraction of its break-up
velocity. It is therefore plausible to assume that the effects can be
observationally interesting and proceed to explore the dynamical processes
that give rise to such tidal interactions. Furthermore, tidal disruption
and collisional stellar mass loss are important channels for supplying mass
to a low-mass MBH (Murphy, Cohn \& Durisen 1991), and so the consequences
of extreme tidal processes may provide observable links between the
properties of the stellar population near the MBH and its evolutionary
history.

At a distance of $\sim\!8$ kpc (Reid 1993), the low mass
$2.6\!\times\!10^6\, M_\odot$ MBH in the Galactic Center (GC) (Ghez et
al. 2000; Sch\"{o}del et al. 2002) is the nearest and observationally most
accessible MBH. Although it is heavily reddened
($A_{K}\!\sim\!3^\mathrm{m}$, Blum et al. 1996), deep high resolution
astrometric, photometric and spectroscopic IR observations of thousands of
stars very close to the MBH provide information on their luminosity,
effective temperature and orbits (e.g. Eckart, Ott \& Genzel 1999; Figer et
al. 2000; Gezari et al. 2002). Since the GC is the obvious first place to
look for evidence of extreme tidal interactions, the results presented here
will be applied to the GC, but it should be emphasized that these physical
mechanisms are generally relevant for MBH in galactic nuclei.




\section{Tidal spin-up by star-star encounters}
\label{s:spinup}

Stars in the high Keplerian velocity field in a dense stellar cusp around a
MBH will suffer numerous hyperbolic tidal encounters over their lifetimes.
Although such encounters transfer some energy and angular momentum from the
orbit to the colliding stars, they rarely remove enough energy for tidal
capture. This is in marked contrast to the situation in the high density
cores of globular clusters without a MBH, where the colliding stars are on
nearly zero-energy orbits and close collisions lead to the formation of
tidal binaries. The effects of hyperbolic encounters on the stars are
mostly transient.  The stellar dynamical and thermal timescales are very
short compared to the mean time between collisions, and so apart from some
mass-loss in very close collisions, the star is largely unaffected. It is
however more difficult for the star to shed the excess angular momentum,
since magnetic breaking operates on timescales of the order of the stellar
lifetime (Gray 1992). High rotation is therefore the longest lasting
dynamical after-effect of a close encounter. Over time, the stellar angular
momentum will grow in a random walk fashion due to successive, randomly
oriented tidal encounters.

We consider the effect of the tides raised by an impactor star of mass $m$
on a target star of mass $\Ms$ and radius $\Rs$ as the impactor follows an
unbound orbit with periapse $r_p$ from the target star. The tilde symbol
will be used below to denote quantities expressed in units where
$G\!=\!\Ms\!=\!\Rs\!=\!1$. In these units, $\widetilde\Omega\!=\!1$ is the
centrifugal break-up angular frequency, $\UE_b\!=\!-1$ is the stellar
binding energy, up to a factor of order unity, and $\Urt\!=\!\Um^{1/3}$ is
the MBH tidal radius.

The orbital energy $\UdE$ and the orbital angular momentum $\UdJ$ that are
transferred from the orbit to the star by an impulsive tidal encounter are
related by $\UdE\!=\!\UWp \UdJ$, where $\UWp$ is the relative angular
velocity at periapse. We assume here for simplicity rigid body rotation.
The change in the stellar angular velocity due to a single parabolic
encounter is then given to the leading order in the linear multipole
expansion by (Press \& Teukolsky 1977)
\begin{equation}
\label{e:dw}
\UdW \simeq \frac{\Um^{2}}{\UI\UWp}\frac{T_2(\eta)}{\Urp^6} \, ,
\end{equation}
where $\UI$ is the stellar moment of inertia (assumed to remain constant),
$T_2$ is the $\ell\!=\!2$ tidal coupling coefficient (calculated
numerically for a given stellar structure model), and
$\eta\!=\![\Urp^3/(1+\Um)]^{1/2}$ is the dimensionless transit time of the
encounter. The steep $\Urp^{-6}$ dependence indicates that most of the
contribution comes from very close encounters, where the linear expansion
no longer holds. The formal divergence at small periapse must be truncated
by non-linear effects, which have to be investigated numerically. Smoothed
Particle Hydrodynamics (SPH) simulations (e.g. Fig.~\ref{f:BHtide}) of
grazing and penetrating encounters show that as $\Urp$ is decreased, $\UdW$
first increases above the value predicted by linear theory, and then
saturates at the onset of mass loss, as the ejecta carry away the excess
angular momentum.

\begin{figure}
\centering
\includegraphics[width=80mm,angle=270]{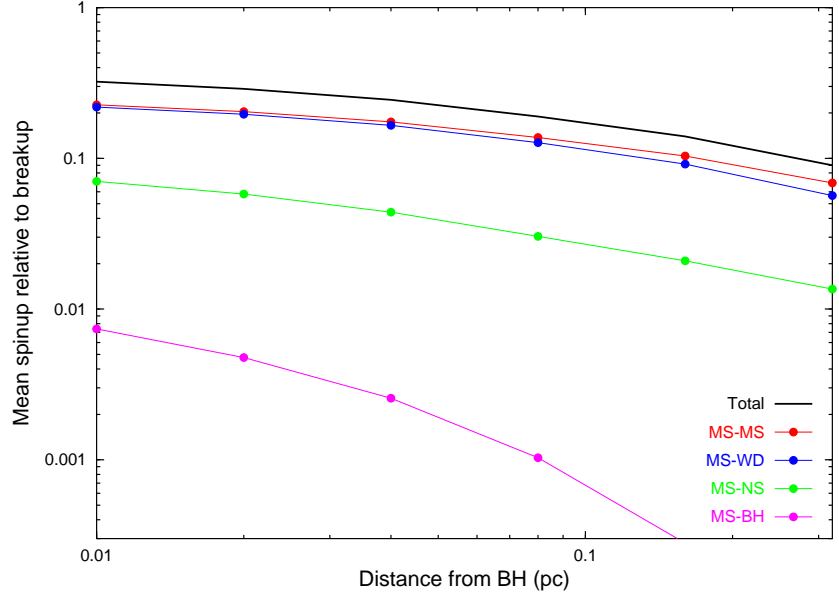}
\caption{
Tidal spin-up in the Galactic Center (Alexander \& Kumar 1999). The total
mean stellar spin-up of a solar type star after 10 Gyr is plotted as
function of distance from the MBH. The separate contribution from
collisions with other MS stars, white dwarfs (WD), neutron stars (NS) and
stellar black holes (BH) are also shown.}
\label{f:spin-up}
\end{figure}

These results can be incorporated into the linear theory by simple
prescriptions, making it possible to calculate the mean spin-up of a test
star by averaging over all impact parameters, orbital energies and impactor
masses for a given model of the nuclear stellar cluster (Alexander \& Kumar
1999). Figure~\ref{f:spin-up} shows the mean spin-up of a solar mass star
after 10 Gyr near the MBH in the GC, as function of distance from the MBH,
in a high density, $n_\star\!\propto\!r^{-1.5}$ cusp of a continuously star
forming population that is deduced to exist there (Alexander \& Sternberg
1999; Alexander 1999). The mean spin-up reaches values as high as
$\sim\!0.3$ within 0.03 pc, ($\sim\!60$ times higher than is typical in the
field) and decreases to $\sim\!10\%$ within 0.3 pc ($\sim\!0.1$ to $0.2$ of
$\Urh$). The spin-up effect falls off only slowly with distance from the
black hole because the increased tidal coupling in slower collisions at
larger distances compensates for the decrease in the stellar density. Thus,
long-lived main sequence stars with inefficient magnetic breaking are
expected to rotate at a significant fraction of their centrifugal breakup
velocity in a large volume of the dense stellar cusp around a MBH .

\section{Tidal scattering by the central black hole}
\label{s:scatter}

Tidal disruption is an important channel for feeding low-mass MBHs that
accrete from a low density cusp where collisional mass-loss is low. Numeric
models of the growth of a MBH in a central cluster suggest that the
fraction of the MBH mass that is supplied by tidal disruption ranges
between $f\!\sim\!0.15$ (Murphy, Cohn \& Durisen 1991) and $f\!\sim\!0.65$
(Freitag \& Benz 2002), depending on model assumptions. We will here
$f\!=\!0.25$ as a representative value.

\begin{figure}
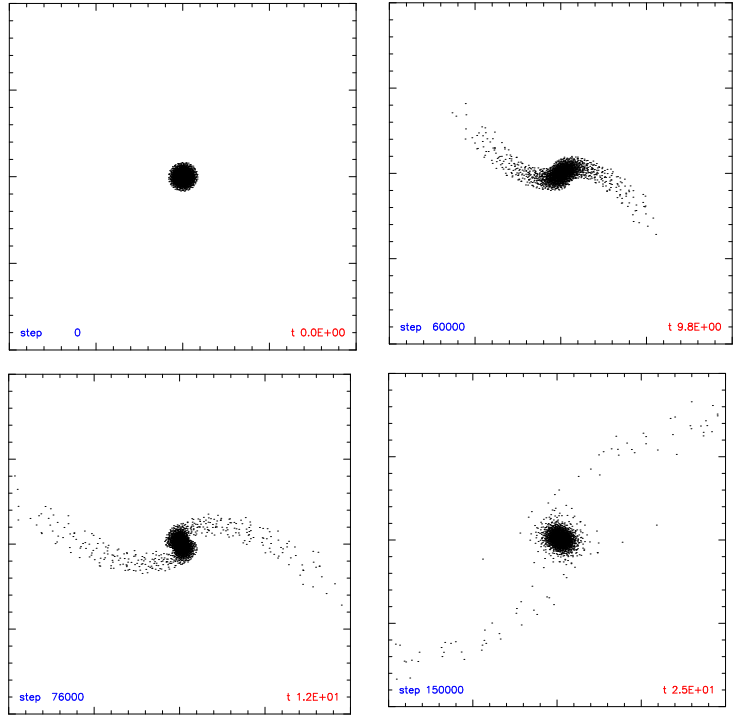


\centering 
\begin{tabular}{cc}
\resizebox*{0.4\textwidth}{!}{\rotatebox{270}{\includegraphics{bh_tide_1.ps}}} &
\resizebox*{0.4\textwidth}{!}{\rotatebox{270}{\includegraphics{bh_tide_121.ps}}} \\
\resizebox*{0.4\textwidth}{!}{\rotatebox{270}{\includegraphics{bh_tide_153.ps}}} &
\resizebox*{0.4\textwidth}{!}{\rotatebox{270}{\includegraphics{bh_tide_301.ps}}} \\
\end{tabular}

\caption{Snapshots from a Smoothed Particle Hydrodynamics (SPH) simulation
of a star undergoing an extreme non-disruptive tidal interaction (``tidal
scattering'') as it passes near a massive black hole. Time is measured in
units of the stellar dynamical timescale. The star passes near the black
hole (located outside the frame) on a parabolic orbit with a peri-distance
1.5 times larger than the tidal disruption distance.  Shortly after
peri-passage ($t\!=\!2$) the star appears to be on the verge of breaking in
two. However, by the end of the simulation, the two fragments coalesce,
leaving a distorted, mixed and rapidly rotating bound object. }
\label{f:BHtide}
\end{figure}

Dynamical analyses of the 2-body scattering by which stars are deflected
into ``loss-cone'' orbits that bring them within $\Urt$ of the MBH (Frank
\& Rees 1976; Frank 1978; Lightman \& Shapiro 1977; Magorrian \& Tremaine
1999; Syer \& Ulmer 1999) show that tidally disrupted stars are typically
on slightly unbound orbits relative to the MBH, and that they mostly
originate from the MBH radius of influence, $\Urh\!\equiv\!\Um/\Us^2$,
where $\Us$ is the 1D velocity dispersion far from the MBH. The stellar
mass enclosed within $\Urh$ is comparable to $\Um$. The cross-section for
such stars to pass within $\Urp$ of the MBH scales as
$\sim\!\Urp$\footnote{This follows from the fact that the stars are on
nearly parabolic orbits and that the scattering is isotropic. First, since
the enclosed stellar mass within $\Urh$ roughly equals the MBH mass, a star
on a plunging orbit will pass the MBH with a velocity slightly above the
local escape velocity. Second, $\Urh$ is where the loss-cone replenishment
efficiency peaks sharply, at the transition from the empty loss-cone
(``diffusive'', or ``small angle scattering'') regime and the full
loss-cone (``pinhole'', or ``large angle scattering'') regime. Deflection
by large angles relative to the loss cone opening angle leads to an
isotropic redistribution of the velocity.}(Hills 1975; see also
\S\ref{s:inspiral}). It then follows that for every star on an orbit with
$0\!\le\!\Urp\!\le\!\Urt$ that is promptly disrupted, there is a star that
skirts the tidal disruption zone on an orbit with
$\Urt\!\le\!\Urp\!\le\!2\Urt$. Such ``tidally scattered'' stars narrowly
escape disruption on their first peri-passage after being subjected to
extreme tidal distortion, spin-up, mixing ad mass-loss
(Fig. \ref{f:BHtide}). We will now argue that there is also a high
probability that these stars will avoid subsequent total disruption, either
by being deflected off their orbit or by missing the MBH due to its
Brownian motion (Alexander \& Livio 2001).

First, the scattering timescale is shorter than the dynamical one, and so
stars wander in and out of the loss-cone several times during one orbital
period. After the first peri-passage, the stars are on very eccentric
orbits with apoastron $\lesssim 2\Urh$, and so there is a considerable
chance that they will be scattered again out of the loss-cone before their
next close passage, and avoid eventual orbital decay and disruption.

Second, the survival probability is further increased by the Brownian
motion of the BH relative to the dynamical center of the stellar system.  A
low mass MBH which evolves in an initially constant density core of radius
$\sim\!\Urh$, is estimated to have Brownian fluctuations with an amplitude
that is much larger than the tidal radius (Bahcall \& Wolf 1976),
\begin{equation} 
\frac{\left\langle\Delta\Ur\right\rangle}{\Urt} \sim
\frac{\Urh}{\Um^{5/6}}\gg 1\,.
\end{equation} 
The Brownian motion proceeds on the dynamical timescale of the core,
which is comparable to the orbital period of the tidally scattered
stars. The orbits of these stars take them outside of $\Urh$, where
they are dynamically affected only by the center of mass of the
nucleus, and not by the relative shift between the MBH and the stellar
mass. Therefore, on re-entry into the volume of influence, their orbit
will not bring them to the same peri-distance from the MBH.

These order of magnitude arguments are verified by more detailed analysis
(see \S\ref{s:inspiral}), which shows that the survival probability of
tidally scattered stars is $P_s\sim\!0.8$--$0.9$. It then follows that the
mass fraction of surviving tidally scattered stars within $\Urh$, which
passed once within $\Urp<2\Urt$ of the MBH, is comparable to the mass
fraction of the MBH supplied by tidal disruption. Depending on the
definition of what constitutes an extreme tidal interaction, there exists a
maximal periapse, parameterized by $b_e\!\equiv\!\Urp/\Urt$, that corresponds to
sufficiently strong tidal interactions. For example, for a solar type star,
$b_e\!=\!1.25$ corresponds to a minimal tidal energy deposition of
$\UdE=0.02$ (Eq. \ref{e:dE}).  Since the angular velocity at periapse is
$\UWp\!=\!(2/b_e^3)^{1/2}\!=\!1.01$ and the solar moment of inertia
$\UI\!=\!0.07$, this corresponds also to a minimal angular spin deposition
of $\UdJ\!=\!0.02$ and a minimal spin-up of $\UdW\!=\!0.28$. Over time, the
mass fraction of tidally scattered stars within $\sim\!\Urh$ will rise to
$f(b_e\!-\!1)P_s\!\sim\!0.05$ (for $f\!=\!0.25$, $b_e\!=\!1.25$ and
$P_s\!=\!0.8$). Tidally scattered stars thus constitute a non-negligible
fraction of the stellar population in the MBH radius of influence, and will
remain there as relics of the early stages of the MBH evolution even after
its mass grows above the tidal disruption limit, possibly detectable by
correlations between unusual spectral properties and highly eccentric orbits.    

\section{Squeezars: Tidally powered stars}
\label{s:squeezar}

A small fraction of the stars that are deflected into orbits with
$\Urp\!\gtrsim\!\Urt$ will be tidally caught by the MBH and spiral into an
ever tighter orbit as the tides gradually extract orbital energy each
peri-passage. The orbital energy that a star has to lose to circularize
from an $\UE\!=\!0$ orbit exceeds its own binding energy by orders of
magnitude,
\begin{equation}
\UE_c = \frac{{\Um}^{2/3}}{2b}\gg1\,,
\end{equation}
where the periapse is parametrized by $b\!\equiv\!\Urp/\Urt$. A tidally
heated star---a ``squeezar''---will ultimately be disrupted by
expanding beyond its Roche lobe or by radiating above its Eddington
luminosity.

The orbital and internal evolution of a tidally heated star in the course
of the inspiral depends on its initial structure and is coupled to the
changes in its mechanical and thermal properties in response to the tidal
heating. One approach to the challenging problem of modeling squeezar
evolution is to consider two simplified cases that likely bracket the range
of possible responses (Alexander \& Morris 1993): (1) Surface heating and
radiative cooling (``hot squeezar''), where the tidal oscillations
dissipate in a very thin surface layer that expands moderately and radiates
at a significantly increased effective temperature (McMillan, McDermott \&
Taam 1987). (2) Bulk heating and adiabatic expansion (``cold squeezar''),
where the tidal oscillations dissipate in the stellar bulk and cause a
large, quasi self-similar expansion at a constant effective temperature
(Podsiadlowski 1996).

Given these prescriptions, the evolution of the squeezar orbit, size,
luminosity and temperature can be derived from the tidal energy deposition
equation
\begin{equation} 
\UdE = \frac{T_2(b^{3/2})\UR^5}{b^6}\,\,\,\,\,\,\,\,\,\,\,\, (\Um\gg1)\,,
\label{e:dE}
\end{equation} 
where $\UR$ is the expanded stellar radius in terms of the original radius,
and the orbital equations for the semi-major axis $\Ua$, the period $\UP$
and eccentricity $e$,
\begin{equation}
\Ua =-\Um /2\UE \,, \,\,\,\,\,\,\,\,\,\,\,\,
\UP = 2\pi \sqrt{\Ua ^{3}/\left(1+\Um \right)}\,, \,\,\,\,\,\,\,\,\,\,\,\,
e=1-\Urp /\Ua \,,
\label{e:orb}
\end{equation}
where $\UE$ is the orbital energy, and Keplerian orbits near the MBH are
assumed.

\begin{figure}
\centering
\includegraphics[width=120mm]{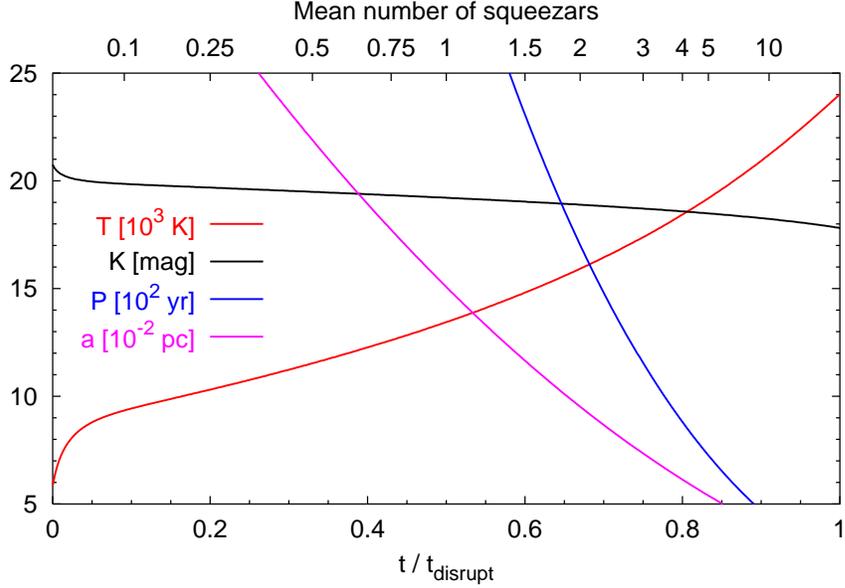}
\caption{The evolution of a hot squeezar in the GC (Alexander \& Morris 2003).
The $1\, M_\odot$ star is on an orbit with $b\!=\!1.5$ and
$P_0\!=\!1.4\! \times \! 10^4$ yr ($t_0\!=\!4.9\! \times \! 10^5$
yr). The star is disrupted when $\UR = b$, at $t_{\rm
disrupt}\!=\!3.7\!\times\! 10^5$ yr. At that point the tidal
luminosity exceeds the intrinsic luminosity by a factor of $\sim\!640$,
but the orbit is still almost parabolic with eccentricity
$e\!=\!1-2.3\!\times\!10^{-4}$. The average of properties of the
leading squeezar as function of the mean number of squeezars can be
read off the top axis.}
\label{f:squeezar}
\end{figure}

Figure~\ref{f:squeezar} shows the evolution of a $1
\Mo$ hot squeezar, with the tidal coupling coefficient calculated for
a model of the Sun (Alexander \& Kumar 2001). In particular, the
squeezar evolutionary model relates the inspiral time $\Ut_0$ to the
initial orbital period $\UP_0$ and the periapse $b$. The mean number
of squeezars at any given time in the GC is given by
$\en\!=\!\overline{t}_0 \Gamma_i$, where $\overline{t}_0$ is the mean
inspiral time, and $\Gamma_i$ is the inspiral event rate (estimated in
\S\ref{s:inspiral}). As shown below, $\en\!\sim\!0.1$--$1$ in the GC if the
loss-cone is replenished by two body scattering in a spherically symmetric
system.

The observational implications of having on average $\en$ squeezars near
the MBH can be expressed by considering the properties of the ``leading
squeezar'' (the one with the shortest period). The leading squeezar has, on
average, completed $\Ut/\Ut_0 = \en/(\en\!+\!1)$ of its
inspiral. Figure~\ref{f:squeezar} also presents the results in terms of the
average properties of the leading squeezar, as function of $\en$. It is
evident that the effects of tidal heating on the leading squeezar can be
quite pronounced even if $\en$ is small, and that the properties of the
leading squeezar can be much more extreme if $\en$ was under-estimated by
the neglect of non-spherical effects.

\section{Prompt disruption .vs. slow inspiral}
\label{s:inspiral}

The orbital decay of a tidally heated star is just one case of a
dissipative interaction that can lead to orbital inspiral. Other
possibilities include gravitational wave (GW) emission (Hils \& Bender
1995; Sigurdsson \& Rees 1997; Freitag 2001, 2003) or drag against a
massive accretion disk (Ostriker 1983; Syer, Clarke \& Rees 1991;
Vilkoviskij \& Czerny 2002). Unlike prompt disruption or tidal scattering,
where the star reaches the MBH directly in less than the initial orbital
period $\UP_0$, slow inspiral proceeds gradually over a timescale
$\Ut_0\!\gg\!\UP_0$, which is typically a steeply rising function of the
periapse. For the extracted orbital energy to power a high luminosity of
gravity waves, tidal heat, or mechanical energy in the disk, as the case
may be, the star has first to decay into a short period orbit. The time
available for inspiral is limited by two-body collisions similar to those
that deflected the star into its eccentric orbit in the first place, since
they can deflect it again to a wider orbit where the dissipation is
inefficient. Because $\Ut_0\!\gg\!\UP_0$, this poses a much more severe
constraint for an inspiraling star than for a promptly disrupted star.

Novikov et al. (1992) estimated that tidal capture by a MBH occurs for
orbits with $b\!<\!b_c\!\sim\!3$\footnote{Based on the requirement that the
orbital energy extracted by the first peri-passage should decrease the
apoapse to $2\Ua\!\le\Ur_0$, where $\Ur_0$ is the distance from which the
star was scattered into the loss-cone. This tidal capture criterion does
not include timescale considerations.}, which means that stars are
scattered into tidal capture orbits and subsequently disrupted at a rate
that is $b_c\!-\!1\!\sim\!2$ times faster than the rate at which they are
scattered into prompt tidal disruption orbits. These considerations led to
the suggestion (Frank \& Rees 1976; Novikov et al. 1992; Magorrian \&
Tremaine 1999) that slow tidal inspiral may be at least as important as
prompt disruption for feeding the MBH and for producing observable tidal
flares (Frank \& Rees 1976). This implies that the already large
contribution of prompt tidally disrupted stars to the mass budget of a
low-mass MBH (\S\ref{s:scatter}) should be further scaled by the ratio of
the cross-sections of tidal capture and prompt disruption, $b_c\!-\!1$. If
the relative contribution of inspiraling stars were indeed so high, the
implications would be far-reaching: stars could supply most or even all of
the MBH mass, thereby establishing a direct link between $\Um$ and stellar
dynamics on a scale of $\Urh$.  However, a small initial periapse does not
in itself guarantee inspiral and ultimate disruption. The star must also
have enough time to complete its orbital decay.

This time constraint can be taken into account correctly by considering
only stars that are scattered from a volume that is close enough to the MBH
so that the inspiral can be completed in time (Alexander
\& Hopman 2003). The inspiral time $\Ut_0$ increases with
$\Urp$. Therefore, the volume from which a scattered star can inspiral
faster than the time it takes for two body relaxation to significantly
deflect it, decreases with $\Urp$. The maximal possible periapse
corresponds to the point where the available volume shrinks to zero (The
truncation of the cusp near the MBH by destructive stellar collisions
also limits the available volume).

Table \ref{t:rates} lists the inspiral event rate, the mean number of
inspiraling stars, and the maximal periapse for inspiral in the GC for
three processes: hot and cold $1\,\Mo$ squeezars and for gravitational wave
inspiral by $0.6\,\Mo$ white dwarfs (WD) that comprise 10\% of an old
stellar population. The rates were estimated for a spherical single-mass
Keplerian stellar cusp, $n_\star\!\propto\!r^{-1.8}$, normalized to contain a
total stellar mass $2.6\!\times\!10^6\,\Mo$ ($=\!m$) within $r_h\!=\!1.8$
pc (after Sch\"{o}del et al. 2002). The predicted prompt disruption rate
for this simple model is $\Gamma_p\!=\!9\times\!10^{-5}\,\mathrm{yr}^{-1}$,
in general agreement with independent estimates from previous studies,
$\Gamma_p\!=\!5\times\!10^{-5}\, \mathrm{yr}^{-1}$ (Syer \& Ulmer 1999) and
$\Gamma_p\!\sim\!\mathrm{few}\times\!10^{-5}\,
\mathrm{yr}^{-1}$ (Alexander 1999). The rate of WD inspiral derived here,
$\Gamma_i\!\sim\!2\!\times\!10^{-7}\,\mathrm{yr}^{-1}$, is also consistent
with the estimates of Sigurdsson \& Rees (1997) and Freitag (2003). The
survival probability of tidally scattered stars can also be calculated
by the formalism of Alexander \& Hopman (2003), and it is found to be close
to unity, $P_s\!\sim0.8$--$0.9$, as anticipated by general arguments
(\S\ref{s:scatter}).

\begin{table}
\caption{Inspiral in the Galactic Center.
\label{t:rates}}
\begin{tabular}{lccc}
Process 	&$\Gamma_i$ (yr$^{-1}$) &$\en$	 	&$\max r_p$	\\
\hline
Hot squeezar	&$3(-6)$		&$0.2$		&$2.1 r_t$	\\
Cold squeezar	&$4(-6)$		&$0.2$		&$2.8 r_t$	\\
\hline
WD gravity waves&$2(-7)$		&$0.04$		&$25 r_s$	\\
\hline
\end{tabular}
\end{table}

However, we find that the {\em tidal} inspiral rate is only $\sim\!0.05$ of
the prompt disruption rate, and {\em not} the factor 2--3 enhancement due
to tidal capture that was assumed by previous studies. We conclude that the
contribution of tidal capture to the MBH mass budget and to the tidal
flaring rate from galactic nuclei is negligible compared to prompt
disruption. Past studies, which assigned similar weights to prompt
disruption and tidal capture, over-estimated the contribution of tidal
disruption to the growth of the MBH by at least a factor of two.

\section{Summary}
\label{s:summary}
We have shown that strong tidal interactions of stars with a MBH or with
other stars in the high density cusp around a MBH can deposit large amounts
of orbital energy and angular momentum in a significant fraction of the
stellar population within a large volume around the MBH. We propose that
such interactions can alter the evolution and appearance of stars in
galactic centers and thereby probe the evolution of the MBH and the stellar
system around it. We explored tidal spin-up by star-star encounters, tidal
scattering by a MBH and tidal inspiral into a MBH. We showed that
tidal capture is inefficient in the presence of two body scattering.

\end{document}